\documentclass{cimento}

\usepackage{graphicx}  

\title{Overview of the experimental quest for the giant pairing vibration.}
\author{M.~Assi\'e}
\instlist{
  \inst{} Universit\'e Paris-Saclay, CNRS/IN2P3, IJCLab, 91405 Orsay, France.}

\begin{document}

\maketitle

\begin{abstract}
The search for the giant pairing vibration (GPV) has a long standing history since the 1970's when it was predicted. First experimental measurements focused on (p,t) transfer reactions in the heavy nuclei and did not show convincing evidence. The discovery of a signal compatible with the GPV in the light carbon isotopes has renewed the interest for the GPV. It triggered new theoretical models showing that the GPV in the heavy nuclei might be too wide or too melted to be observed and triggered new experiments with radioactive probes based on ($^{6}$He,$^{4}$He) transfer.
\end{abstract}

Nuclei in interaction with external fields display a wide variety of collective vibrations known as giant resonances, associated with various degrees of freedom and multipolarities. The giant isovector dipole resonance and the giant isoscalar quadrupole resonance are among the most studied ones. Through general symmetry arguments relating particles to holes, a particular mode associated with vibrations in the number of particles was predicted in the 1970s \cite{Bro77}, the Giant pairing vibration (GPV). It is due to the collective superposition of many particle-particle states, in analogy with the Giant (dipole, quadrupole) resonances that are due to the collective superposition of many particle-hole states. The analogy between particle-hole (shape) and particle-particle (pairing) excitations became well established and thoroughly explored by Broglia and co-workers \cite{Bro73}. 
However, although the first experimental evidence for the isovector giant dipole resonance dates back to 1937, no clear evidence for the existence of the giant pairing vibration was ever found in the heavy nuclei. This topic has regained interest since the recent work of Cappuzzello and coworkers \cite{Cap15} who identified a signal compatible with the GPV in the light carbon isostopes. In this paper, we will review the experimental search for the GPV. A full review of the theoretical and experimental search for GPV can be found in ref.\cite{Ass19} and a review on the experimental work in the light nuclei in ref.\cite{Cav19}.

\section{How to probe the GPV ?}

The collective character of the particle-hole (surface) vibration is probed by inelastic scattering reactions. In the same fashion two-particle transfer reactions provide much of our knowledge of pairing correlations.  For excitations to 0$^+$ states these reactions are important probes of collective pairing excitations in nuclei. This has the same origin as the collectivity of surface vibrations in inelastic scattering. Namely all configurations contribute with the same phase to the two-particle transfer form factor leading to the collective pairing state (a vibration in gauge space).
The existence of pair correlations is known to provide an enhancement in the magnitude of the ground-state to ground-state transition matrix elements between systems that differ by a number of two nucleons. Analogous enhancements are expected from particle-particle correlations involving transitions to higher single-particle shells.

The analogy between shape and pairing can be taken further. Near closed shell, nuclei with two identical particles added or removed from a closed-shell configuration should be close to a quantum fluid limit, since the pairing correlations are not strong enough to overcome the large single-particle energy required to add a pair. 
There, strongly enhanced L=0 transitions manifest themselves following a vibrational pattern (similar to vibrations in shape), in which transferring a pair of nucleons changes the number of phonons by one. Low-lying pair-vibrational structures have been observed around $^{208}$Pb by using conventional pair-transfer reactions such as (p,t) and (t,p) \cite{Boh66}.
Nuclei with many particles (pair quanta) outside of a closed-shell configuration (i.e. at the middle of a shell) correspond to a superconducting limit, where there is a static deformation of the pair field and rotational behavior results (similar to shape rotation). Ground-state to ground-state transition is observed between monopole states and follows a rotational scheme. A textbook example would be the pair-rotational sequence comprising the ground states of the even-even Sn isotopes around $^{116}$Sn \cite{Bri05}. 

Taking the analogy even further, it has long been predicted that there should be a concentration of strength, with L = 0 character, in the high-energy region (10– 15 MeV) of the pair-transfer spectrum. This is called the Giant Pairing Vibration (GPV) and is understood micro-scopically as the coherent superposition of 2p (addition mode) or 2h (removal mode) states in the next major shell 2$\hbar\omega$ above (below) the Fermi surface. Similar to the well-known pairing vibrational mode (PV) \cite{Bro77,Bes66,Boh98}, which involves spin-zero-coupled pair excitations across a single major shell gap. 

Thus, pairing vibrations are evidenced through two-particle transfer reactions. They manifest themselves as a L=0 transition mode from an A nucleus to a A$\pm$2 nucleus. They are expected to lead to a large bump in the two-neutron transfer energy spectrum.   Various independent theoretical calculations converge in predicting the GPV as a strong mode typically located around 70/A$^{1/3}$ MeV in two-neutron L=0 transfer channel with a width of 7.8/ A$^{1/3}$ MeV and carrying a cross-section which is 20\%-100\% of the ground state one \cite{Bro73,Oer01,For02}. The study of the GPV would also provide crucial information on the pairing interaction:  the transfer cross-section depends on the form-factor of the two transferred neutrons. It has been shown that this form factor corresponds to the perturbation of the pairing field during the excitation of the system \cite{Kha04,Oer01}.  

Intuitively, the collectivity of the GPV should increase with the mass of the nucleus. Therefore, in a simple picture, its strength is expected to be maximum for the heaviest nuclei, such as Sn and Pb isotopes, where numerous nucleons may contribute coherently. Two candidates have been envisaged the ”normal” nuclei (closed shells), like Pb \cite{Her85}, or the superfluid nucleus (mid-shell), namely Sn, where more pairs can contribute.  Experimental investigations of the GPV focussed on simple probes like (p,t) and (t,p) with various conditions.

\section{Review of the experimental search for GPV}
\subsection{In the heavy nuclei}
In the 60‘s and 70‘s, the searches for the GPV have focused on (p,t) reactions at high energy for both Pb and Sn isotopes. However it remained unsuccessful. There could be several reasons as mentionned in \cite{Mou11} :
\begin{itemize}
\item The L matching conditions are an of great importance. The proton incident energy should be high enough to excite a 14 MeV mode but not too high in order not to hinder the L=0 transfer. The smaller the proton energy the larger the cross section for L=0 modes.
\item The use of a spectrometer is decisive in order to precisely measure the triton in the exit channel. The only reported search for the GPV with Ep $\approx$ 50 MeV used Si detectors, and was plagued by a strong background \cite{Cra77}.
\item As the L = 0 cross sections are known to exponentially increase when approaching 0 degree, the measurement has to be performed at small angles and is even better if it includes 0 degree.
\end{itemize}

There was a revival of the experimental GPV search in the 2000’s with several experiments aiming at improving the three experimental conditions mentioned above. All used a spectrometer for the triton measurement to im- prove the measurement at 0 degree. Several attempts with different proton energies were performed.
The first attempt used a 60 MeV proton beam produced at the iThemba LABS facility in South Africa impinging on $^{208}$Pb and $^{120}$Sn targets respectively \cite{Mou11}. No evidence for the GPV was found in the region of interest for both targets.

\begin{figure}
\center\resizebox{0.7\textwidth}{!}{ \includegraphics{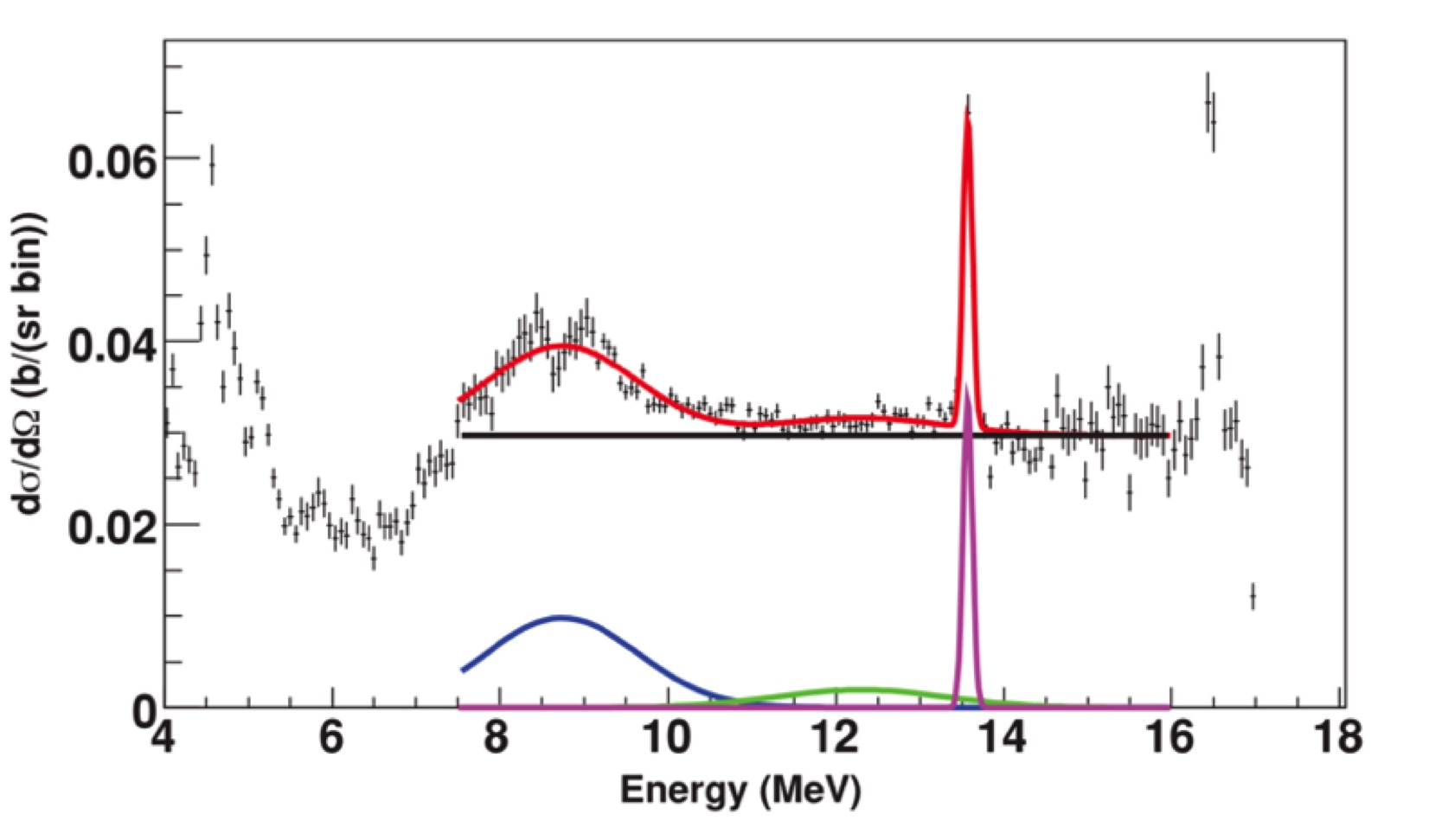}}
\caption{(Color online)  Excitation energy spectrum of $^{118}$Sn for the 0 degree measurement at Ep = 50 MeV. The linear background, deep-hole states (around 9 MeV), fit for a possible GPV (around 12 MeV), oxygen contaminant (around 13.5 MeV) and total fitting function are shown. The bin width is 67 keV/bin. From  \cite{Mou11}.\label{fig_GPV_iThemba}}
\end{figure}

The measurement was repeated with a 50 MeV and a 60 MeV proton beam and the K = 600 QDD magnetic spectrometer in zero degree mode to combine the best experimental conditions to probe the GPV. A strong proton background with a rate 500 times higher than that of the tritons of interest was produced by protons scattering off the beam stop . The tritons were identified by time-of-flight and it removed most of the background. The excitation energy spectrum obtained for $^{118}$Sn is shown in Fig. \ref{fig_GPV_iThemba}. The deep holes contribution between 8 and 10 MeV is stronger in the 0 degree spectrum than at 7 degree indicating a possible low L composition of this background. A fit of the different component assuming a width between 600 keV and 1 MeV for the GPV was performed. It leads to a higher limit on the cross-section for populating the GPV between 0.13 and 0.19 mb over the angular acceptance of the spectrometer ($\pm$2 degrees).

\begin{figure}
\center\resizebox{0.7\textwidth}{!}{ \includegraphics{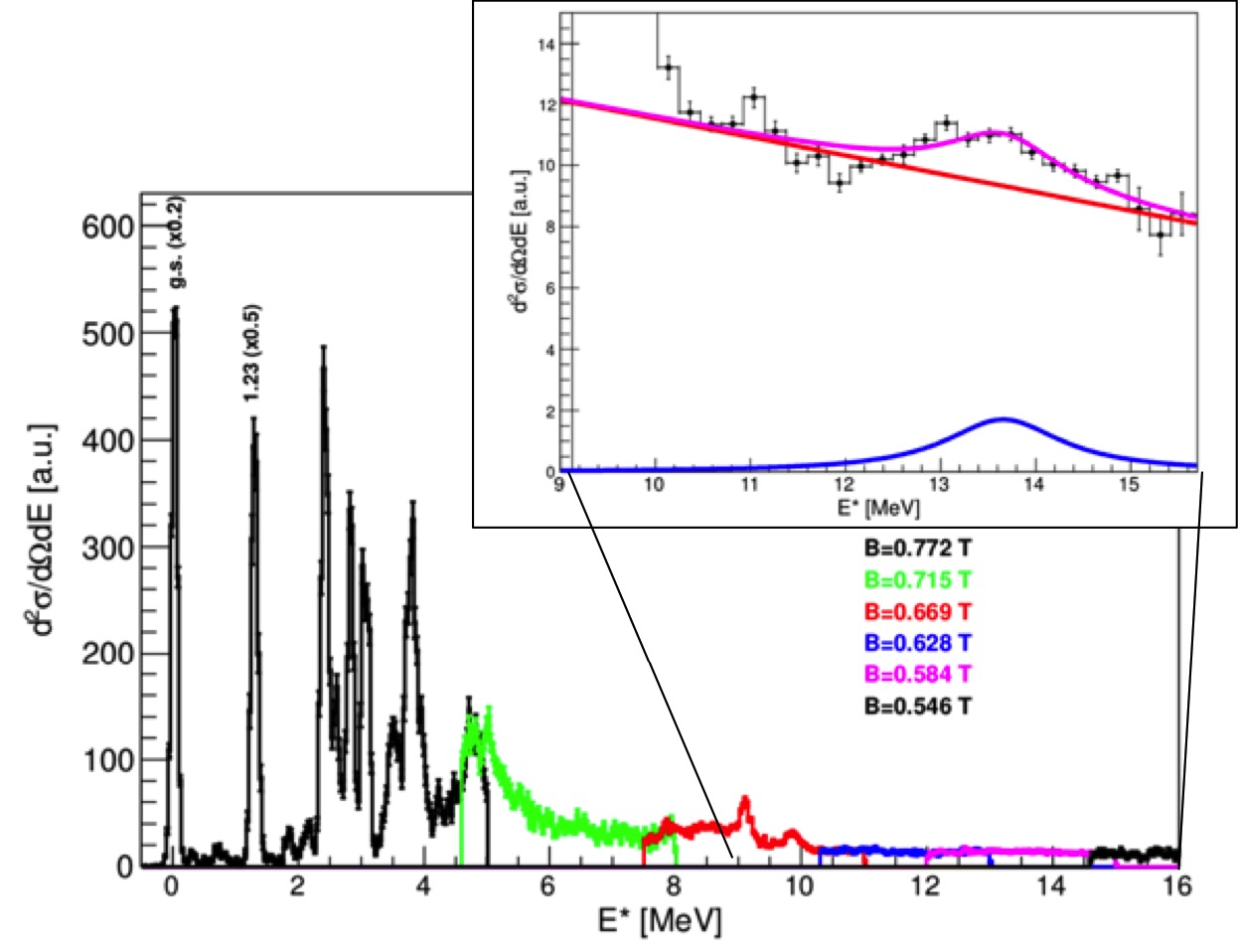}}
\caption{(Color online)  (Color online) Excitation energy spectrum of $^{118}$Sn for the six spectrometer settings at Ep = 35 MeV. A zoom on the GPV region is shown in the insert where a lorentzian fit of the GPV was performed. From  \cite{deN14}.\label{fig_GPV_Catania}}
\end{figure}

The last attempt with (p,t) reaction was performed at LNS Catania with a proton beam produced by the cyclotron accelerator at Ep = 35 MeV impinging on a $^{120}$Sn target \cite{deN14}. The lower proton energy was supposed to enhance the L=0 cross-sections and favor the population of the GPV. The measurement was performed with the MAGNEX spectrometer which allows to cover a range of about 7 MeV in the expected GPV energy region. The excitation energy function obtained for $^{118}$Sn is shown in Fig.\ref{fig_GPV_Catania} for the six magnetic settings of the spectrometer. The tritons were identified from their energy loss as a function of their position in the focal plane so that very small background contribution remains. The spectrum zoomed in the region of interest for the GPV shows a small bump over the background in the same energy region as the previous measurements at 50 and 60 MeV. The width was fitted to 1.5 $\pm$ 0.4 MeV.

No clear evidence for a GPV mode has been found from the searches through (p,t) reactions. (t,p) transfer reactions should also be investigated to rule out any difference between two-neutron stripping and two-neutron pick-up reactions.

\subsection{In the light nuclei}

The search for the GPV in the heavy nuclei was almost stopped when ref.\cite{Cap15} revealed evidences for a GPV in the light carbon isotopes. The GPVs are observed at about 17 MeV and 13.7 MeV in $^{14}$C and $^{15}$C respectively with widths of the order of 1 to 1.5 MeV. The cross-sections at 84 MeV are of the order of 0.3 and 0.4 mbarn respectively. The angular distribution for the GPV state was also extracted and has a L=0 character as expected. 
The author claim that the reaction mechanism plays an important role in populating the GPV. The two-nucleon transfer reaction ($^{18}$O,$^{16}$O) is better suited because:
\begin{itemize}
\item it is better matched in Q-value for L=0 transfer following Brink's conditions
\item the survival of a preformed pair in a transfer process is favored when the initial and final orbitals are the same
\end{itemize}.

The authors further confirmed their results with the study of the two-neutron decay of the GPV and the observation of another GPV in $^{11}$Be with the same transfer reaction. They also investigated the same reactions as in the original paper but at higher energy (275 MeV against 84 MeV in the first measurements) and confirmed their results.

\begin{figure}
\center\resizebox{1.0\textwidth}{!}{\includegraphics{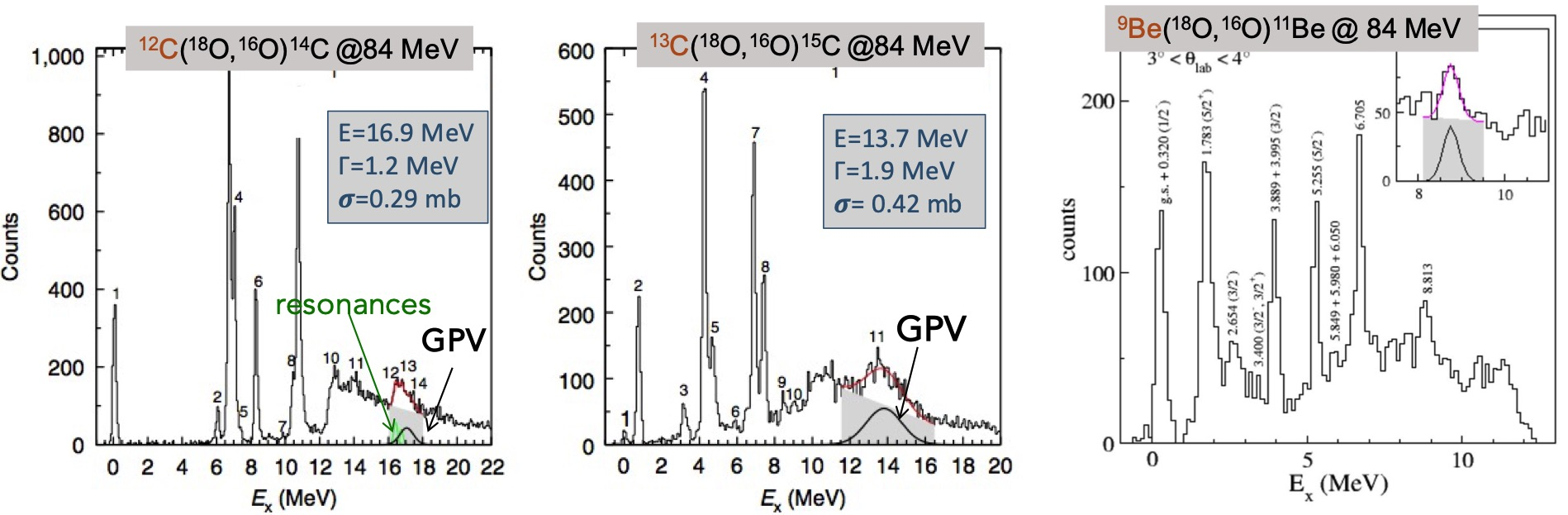}     }
\caption{(Color online) Excitation energy spectra for the $^{14}$C, $^{15}$C and $^{11}$Be populated through the two-nucleon transfer reaction ($^{18}$O,$^{16}$O), extracted from refs.\cite{Cap15,Cav19}.}
\end{figure}

\section{Where do we stand today ?}

 The non-observation of the GPV in the heavy nuclei, when all pairing theories predict it, is a puzzle. Today, we stand at a crossroad : either it is due to the reaction mechanism of the (p,t) transfer reaction that is not well suited for this kind of study or the GPV cannot be observed in the heavier nuclei because it is too wide to be observed.
 The arguments in favor of the latter point are many : the density of states in the heavy nuclei is maybe too high to observe the GPV that is melted with other contributions. Laskin and collaborators \cite{Las16} suggested that as the main contribution to L=0 transfer comes from low-l orbitals (mainly the s$_{1/2}$ orbital) for (p,t), they have low centrifugal barrier and thus, they acquire large widths, so that the GPV would be too wide to be observed.

As for the reaction mechanism aspect, new attempts to search for the GPV in heavier nuclei have been undertaken recently. Several theoritical work \cite{For02,Das15} pointed out that pair transfer reactions using weakly bound projectile would be more suited in terms of Q-value and transferred angular momentum matching, like ($^{6}$He,$^{4}$He) transfer.

\subsection{ Search for GPV through ($^{6}$He,$^{4}$He) reactions}

 The $^{208}$Pb($^{6}$He,$\alpha$) reaction has been investigated at GANIL \cite{Ass09} with the $^{6}$He beam produced by the Spiral1 facility at 20 A MeV with an intensity of 10$^{7}$ pps. The detection system was composed of an annular Silicon detector. The background was very important, due to the various channels of two-neutron emission from $^{6}$He (namely breakup) and also to the channeling in the detector of the elastically scattered $^{6}$He beam. No indication of GPV was found in this experiment.

Another experiment has been performed at TRIUMF with the IRIS set-up\cite{iris} to investigate $^{116}$Sn($^6$He,$^4$He) and $^{116}$Sn($^{18}$O,$^{16}$O) at 8 MeV/u (spokesperson : R.M. Clark and A.O. Macchiavelli). The preliminary results reflect the difficulties in using Silicon/CsI telescopes to investigate pair transfer due to the large background from reactions in the CsI, channeling and inter-strip effects in the Silicon.

\section{Conclusion}
The search for the GPV has triggered a lot of experimental efforts, combined with theoretical work to improve the predictions and interpret its non-observation. However, the quest for the GPV in heavy nuclei has not come to an end : further investigations with spectrometers instead of particle arrays and further searches of other types of GPV like the proton-proton GPV is still in his infancy.

\end{document}